\documentclass[article, twocolumn,
   aps, pra,
  amsmath,amssymb,
  longbibliography,
  ]{revtex4-1}
\usepackage{graphicx,color}
\usepackage{amsmath}
\usepackage{natbib}
\usepackage{epsfig}
\begin{document}

\title{Note: Shoving model and the glass transition in one-component plasma}

\author{Sergey Khrapak}\email{Sergey.Khrapak@gmx.de}
\affiliation{Joint Institute for High Temperatures, Russian Academy of Sciences, 125412 Moscow, Russia}

\begin{abstract}
A modified shoving model is applied to estimate the location of the glass transition in a one-component plasma. The estimated value of the coupling parameter $\Gamma\simeq 570$ at the glass transition is compared with other predictions available in the literature.
\end{abstract}

\date{\today}

\maketitle

A one-component plasma (OCP) model is an idealized system of point charges immersed in a uniform neutralizing background of opposite charge~\cite{BrushJCP1966,BausPR1980,IchimaruRMP1982}. This
model is of relevance in a wide interdisciplinary context, including laboratory and space plasmas, planetary interiors, white dwarfs, liquid metals, and electrolytes. There are strong relations to various soft matter systems such as charged colloidal suspensions and complex (dusty) plasmas~\cite{FortovPR,FortovBook,IvlevBook,ChaudhuriSM2011,BeckersPoP2023}. OCP represents a very convenient system to test and verify the applicability of different theoretical approaches used in condensed matter research.   

The mobile particles forming the OCP are interacting via a very soft and long-ranged Coulomb repulsive interaction potential, $\phi(r)= e^2/r$, where $e$ is the electric charge and $r$ is the distance between a pair of particles. The particle-particle correlations and thermodynamics of the OCP are characterized by a single dimensionless coupling parameter $\Gamma=e^2/aT$, where $a=(4\pi n/3)^{-1/3}$ is the Wigner-Seitz radius in three dimensions (3D), and $T$ is the temperature in energy units ($\equiv k_{\rm B}T$). The coupling parameter is equivalent to the inverse temperature in conventional matter. Since the interaction potential is purely repulsive, the OCP does not exhibit a gas-liquid phase transition, gas-liquid coexistence, critical and gas-liquid-solid triple points. The one-dimensional phase diagram of the OCP is very simple. There is a phase transition from the fluid phase to the body-centred cubic (bcc) solid phase at sufficiently strong coupling (low temperature), $\Gamma_{\rm fr}\simeq 174$~\cite{DubinRMP1999,KhrapakCPP2016}, where the subscript ``fr'' refers to freezing. There is also a gas-to-liquid dynamical crossover, which has been recently located at $\Gamma/\Gamma_{\rm fr}\simeq 0.05$, that is at $\Gamma\sim 10$~\cite{HuangPRR2023,Feng2024}.   

Over the years there have been predictions that supercooled OCP fluid might exhibit a glass transition at strong coupling. However, the location of this transition differs greatly in these studies. Ichimaru and Tanaka used a generalized viscoelastic theory to demonstrated the possibility of a glass transition at $\Gamma_{\rm g} = 900 - 1000$~\cite{IchimaruPRL1986}. Cardenas and Tosi~\cite{CardenasPhysB2004} studied the supercooled-fluid region and the transition to an amorphous glassy state in the OCP within the replica-symmetry-breaking scenario developed by Franz and Parisi~\cite{FranzPRL1997}. They obtained the threshold value $\Gamma_{\rm g}\simeq 1500$ for the glass transition. Glass transition properties for the Yukawa (screened Coulomb) potential were investigated by Yazdi {\it et al.} using the traditional mode coupling theory (MCT) with the structural information obtained from the Ornstein-Zernike relation and the hypernetted-chain (HNC) approximation closure~\cite{YazdiPRE2014}. In the infinite screening length limit, corresponding to the OCP, they obtained $\Gamma_{\rm g}\simeq 590$. Lucco Castello and Tolias combined the traditional MCT with three closures of the integral equation theory of liquids to estimate the glass transition line in the Yukawa system~\cite{CastelloMolecules2021}. With HNC closure they located the glass transition point in the OCP at $\Gamma_{\rm g}\simeq 575$, slightly lower than the result of Yazdi {\it et al.} With the isomorph-based empirically-modified hypernetted chain (IEMHNC) approach~\cite{ToliasPoP2019} and the variational modified hypernetted chain (VMHNC) approximation~\cite{Rosenfeld1986}, the obtained coupling parameters are almost a factor of two lower,  $\Gamma_{\rm g}\simeq 290$ and  $\Gamma_{\rm g}\simeq 280$, respectively.

Considerable variation in the locations of the glass transition point reported in different studies suggests checking these predictions with other methods and tools. The purpose of this Note is to apply a simple model of the glass transition based on elastic arguments -- the shoving model~\cite{DyrePRB1996,DyrePRE2004}. The model considers a ``flow event'', which requires a local volume increase. The activation energy for a flow event is identified with the work done in shoving
aside the surrounding liquid. This work can be expressed using the infinite frequency elastic moduli (infinite frequency shear modulus in the original formulation~\cite{DyrePRB1996}). The shoving model is one of the elastic models discussed in connection to glass-forming liquids~\cite{DyreRMP2006}. In the simplest approximation these elastic models lead to a Lindemann-like criterion for the glass transition~\cite{DyreRMP2006}, an analogy that will be elaborated further below. At the moment we remind that the energy-landscape version of the shoving model predicts that the following dimensionless combination is nearly constant at the glass transition~\cite{DyreJCP2012}:
\begin{equation}\label{shoving1}
\frac{G_{\infty}}{nT}\frac{K_{\infty}+4G_{\infty}/3}{2K_{\infty}+11G_{\infty}/3}\simeq {\rm const}.
\end{equation}                  
Here $K_{\infty}$ and $G_{\infty}$ are the infinite frequency bulk and shear moduli and no distinction between the high-frequency plateau moduli and idealized instantaneous affine moduli (see Ref.~\cite{DyreJCP2012} for details) is made in case of the OCP.    
The value of the constant in Eq.~(\ref{shoving1}) is expected to be quasi-universal. For a wide range of metallic glasses the values of the const are scattered in the vicinity of ${\rm const}\simeq 30$. Actually, in Fig.~2 of Ref.~\cite{DyreJCP2012} the data points are scattered around the value of $\simeq 0.03$, but the results are expressed in units of GPa cm$^3$/J = $10^3$.      

The instantaneous bulk modulus of the OCP system is infinite: Due to very soft and long-ranged character of the Coulomb interaction, the dispersion relation has a plasmon dispersion instead of the conventional acoustic one. If we formally substitute $K_{\infty}\rightarrow \infty$ in the Eq.~(\ref{shoving1}) and adopt the constant appropriate for metallic glasses, we obtain $\tilde{G}_{\infty}\simeq 60$ as a preliminary estimate of the glass transition point in the OCP. Here $\tilde{G}_{\infty}=G_{\infty}/nT$ is the reduced shear modulus. Within the quasi-localized charge approximation (QLCA)~\cite{GoldenPoP2000}, the elastic moduli of the OCP system can be directly expressed via the excess internal energy. At strong coupling, the OCP excess energy is not very sensitive to whether OCP forms a fluid, solid, or glass. The ion sphere model provides a reasonable estimate~\cite{BausPR1980,IchimaruRMP1982}, resulting
in~\cite{KhrapakMolecules12_2021,KhrapakPoP10_2019}     
\begin{equation}
\tilde{G}_{\infty}\simeq 0.1936 \frac{e^2 n^{1/3}}{T}\simeq 0.12\Gamma.
\end{equation}
In terms of the coupling parameter this yields $\Gamma_{\rm g}\simeq 500$. Still, this estimate is not very convincing, because it neglects the contribution from the longitudinal collective mode. 

Equation~(\ref{shoving1}) should be modified to account for the non-acoustic character of the longitudinal collective mode in the OCP system. We discuss this modification below. Let us start with estimating the amplitude of the atomic vibrations in the harmonic approximation. We have
\begin{equation}\label{amplitude}
\left\langle \delta r^2 \right\rangle=\frac{3T}{m}\left\langle \frac{1}{\omega^2}\right\rangle,
\end{equation}      
where $m$ is the atomic mass. Averaging can be performed over normal modes,
\begin{equation}
\left\langle \frac{1}{\omega^{2}}\right\rangle = \frac{1}{3N}\sum_{\rm k}\omega_{\bf k}^{-2}.
\end{equation}  
Furthermore, the sum over frequencies can be converted to an integral over $k$ using the standard procedure
\begin{equation}
\frac{1}{V}\sum_{\bf k}(...)\rightarrow \frac{1}{(2\pi)^3}\int (...) d{\bf k},
\end{equation}
where $V$ is the volume. One longitudinal (compressional) and two transverse (shear) modes are supported in the solid state. We get
\begin{equation}\label{kaverage}
\left\langle\frac{1}{\omega^2}\right\rangle = \frac{1}{6\pi^2 n}\int_0^{k_{\rm max}}k^2dk\left(\frac{1}{\omega_l^2}+\frac{2}{\omega_t^2}\right),
\end{equation} 
where $\omega_l$ and $\omega_t$ are the frequencies of the longitudinal and transverse modes and the cutoff $k_{\rm max}= (6\pi^2 n)^{1/3}$ ensures that $\langle {\mathcal X} \rangle = {\mathcal X}$ for a quantity ${\mathcal X}$, which is independent of $k$.   
Substituting the acoustic dispersion relations $\omega_l=c_lk$ and $\omega_t=c_tk$ into \mbox{Eq.~(\ref{kaverage})} and taking into account the relations between the sound velocities and elastic moduli $mnc_l^2=M_{\infty}=K_{\infty}+\tfrac{4}{3}G_{\infty}$ and $mnc_t^2=G_{\infty}$, we get 
\begin{equation}\label{Lindemann}
\frac{\langle \delta r^2\rangle}{\Delta^2}=\frac{3nT}{(6\pi^2)^{2/3}G_{\infty}}\frac{2K_{\infty}+11 G_{\infty}/3}{K_{\infty}+4G_{\infty}/3}.
\end{equation}
Here $\Delta=n^{-1/3}$ is the average interatomic separation and Eq.~(\ref{Lindemann}) represents an expression for the Lindemann measure in the harmonic approximation. According to the Lindemann melting rule, melting of a crystal occurs when the average vibrational amplitude exceeds a universal fraction ($\sim 0.1$) of the inter-atomic distance. By virtue of Eq.~(\ref{shoving1}), an analogue of the Lindemann rule applies to the glass transition, as discussed previously~\cite{DyreRMP2006}. The only difference is in the relative vibrational amplitude. Adopting ${\rm const}\simeq 30$ in Eq.~(\ref{shoving1}) we arrive at 
\begin{equation}\label{Lindemann1}
\langle \delta r^2\rangle/\Delta^2\simeq 0.0066
\end{equation}
at the glass transition point. 

It is now obvious how the shoving approximation should be modified in case of the OCP. Performing averaging in Eq.~(\ref{kaverage}) we have to substitute the plasmon dispersion $\omega_l(k)\simeq \omega_{\rm p}$ instead of the acoustic one. Here $\omega_{\rm p}=\sqrt{4\pi e^2n/m}$ is the plasma frequency. The acoustic approximation for the transverse dispersion relation remains adequate. Even more accurate results can be expected by substituting almost exact QLCA long-wavelength dispersion relations for $\omega_l(k)$ and $\omega_t(k)$~\cite{KhrapakPoP2016,KhrapakIEEE2018} in Eq.~(\ref{kaverage}). There is no need to perform averaging here. This has been already done in connection to the self-diffusion mechanism and the Stokes-Einstein relation between the self-diffusion and viscosity coefficients in the strongly coupled OCP~\cite{KhrapakMolecules12_2021}. The result, which applies equally to strongly coupled fluid and amorphous solid phases is $\langle \omega_{\rm p}^2/\omega^2\rangle\simeq 9.76$~\cite{KhrapakMolecules12_2021,KhrapakPhysRep2024}. Combining Eqs.~(\ref{amplitude}) and (\ref{Lindemann1}) we get the following condition for the glass transition in the OCP model
\begin{equation}
\Gamma_{\rm g}\simeq 570.
\end{equation}    
This is our best estimate for the glass transition point within the shoving model. It should be pointed out, however, that the shoving model is an approximation and the value of the constant in Eq~(\ref{shoving1}) based on metallic glasses may not be optimal for the OCP system. Nevertheless, the present estimate correlates much better with recent MCT predictions of Refs.~\cite{YazdiPRE2014,CastelloMolecules2021}, compared to earlier works~\cite{IchimaruPRL1986,CardenasPhysB2004}.

To summarize, the energy-landscape version of the shoving model has been modified to account for the non-acoustic character of the longitudinal collective mode in the OCP system. This modified version of the shoving model predicts the glass transition in OCP at $\Gamma_{\rm g}\simeq 570$. Similar modifications can be straightforwardly implemented in other simple fluids for which deviations from acoustic asymptotes are important. This includes for instance 2D OCP as well as 2D and 3D Yukawa fluids and related systems with soft interaction potentials~\cite{KhrapakPoP2017}. Generalization of the shoving model to mixtures seems desirable in view of applications to complex plasma experiments targeting the glassy state with polydisperse dust.    



\bibliography{SE_Ref}

\end{document}